\documentclass[%
reprint,
superscriptaddress,
amsmath,amssymb,amsthm,
aps,
soul
]{revtex4-2}

\usepackage{hyperref}
\hypersetup{
  colorlinks   = true, 
  urlcolor     = blue, 
  linkcolor    = blue, 
  citecolor   = blue 
}
\usepackage{algorithm2e}
\SetKwComment{Comment}{/* }{ */}
\RestyleAlgo{ruled}

\usepackage{graphicx}
\usepackage{dcolumn}
\usepackage{bm,soul}
\usepackage{xcolor}
\usepackage{braket}

\usepackage{color}

\begin{document}

\preprint{APS/123-QED}

\title{Quantum-Train: Rethinking Hybrid Quantum-Classical Machine Learning in the Model Compression Perspective}

\author{Chen-Yu Liu}
\email{d10245003@g.ntu.edu.tw}

\affiliation{Graduate Institute of Applied Physics, National Taiwan University, Taipei, Taiwan}
\affiliation{Foxconn Research, Taipei, Taiwan}

\author{En-Jui Kuo}
\email{kuoenjui@umd.edu}
\affiliation{Foxconn Research, Taipei, Taiwan}
\affiliation{Physics Division, National Center for Theoretical Sciences, Taipei, Taiwan}

\author{Chu-Hsuan Abraham Lin}
\affiliation{Foxconn Research, Taipei, Taiwan}
\affiliation{Department of Electrical and Electronic Engineering, Imperial College London, London, UK}

\author{Jason Gemsun Young}
\affiliation{Industrial Technology Research Institute,  
Taipei, Taiwan}

\author{Yeong-Jar Chang}
\affiliation{Industrial Technology Research Institute,  
Taipei, Taiwan}

\author{Min-Hsiu Hsieh}
\email{min-hsiu.hsieh@foxconn.com}
\affiliation{Foxconn Research, Taipei, Taiwan}

\author{Hsi-Sheng Goan}
\email{goan@phys.ntu.edu.tw}
\affiliation{Graduate Institute of Applied Physics, National Taiwan University, Taipei, Taiwan}
\affiliation{Department of Physics and Center for Theoretical Physics, National Taiwan University, Taipei, Taiwan}
\affiliation{Center for Quantum Science and Engineering, National Taiwan University, Taipei, Taiwan}
\affiliation{Physics Division, National Center for Theoretical Sciences, Taipei, Taiwan}


\begin{abstract}
We introduces the Quantum-Train(QT) framework, a novel approach that integrates quantum computing with classical machine learning algorithms to address significant challenges in data encoding, model compression, and inference hardware requirements. Even with a slight decrease in accuracy, QT achieves remarkable results by employing a quantum neural network alongside a classical mapping model, which significantly reduces the parameter count from $M$ to $O(\text{polylog} (M))$ during training. Our experiments demonstrate QT's effectiveness in classification tasks, offering insights into its potential to revolutionize machine learning by leveraging quantum computational advantages. This approach not only improves model efficiency but also reduces generalization errors, showcasing QT's potential across various machine learning applications.


\end{abstract}

\maketitle

\section{Introduction}

Whilst machine learning (ML) has seen huge success in recent years \cite{carrasquilla2017machine,wetzel2017unsupervised,van2017learning,van2018learning,schindler2017probing, PhysRevResearch.4.043031,seif2019machine,PhysRevB.105.235136,kuo2021quantum, rsnn1, Tsai_2020, Tsai_2022}, its quantum counterpart quantum machine learning (QML) is also in rapid development. QML represents a groundbreaking intersection that leverages the unparalleled computational powers of quantum mechanics to transform neural network training and functionality. Noticeably, quantum neural networks (QNNs), through quantum superposition and entanglement, are able to evaluate multiple outcomes and results simultaneously. This could theoretically accelerate the training and learning process \cite{qml1, qml2, qml3, qml4}. Besides QNNs, the quantum kernel method stands out by employing quantum operations to craft kernels through the inner products of quantum states. These kernels are subsequently applied to classical data that has been transformed into the quantum realm \cite{qml5}. Additionally, integrating Grover's search algorithm with QML for classification tasks offers potential enhancements \cite{gs1, gsml1}. QML harbours tremendous potential as a powerful instrument for deciphering complex data sets, poised to drive revolutionary changes across diverse domains. QML's applications are wide-ranging and impactful, including breakthroughs in drug discovery, large-scale stellar classification, natural language processing, recommendation systems, and generative learning models \cite{qmlapp1, qmlapp2, qmlapp3, qmlapp4, qmlapp5, qnlp1, qrs1, qrs2, qirs1, fsqc1, qgan1, qgan2, qgan3, qgan4, qgan5, hu2024sparse}. 

Despite its promising advantages and future, QML is still in its early development stages, due to numerous obstacles that must be overcome to realize its full potential and practicality. Critical challenges involve tackling the learnability \cite{qnnlearn1, qnnlearn2, qnnlearn3, qnnlearn4, qnnlearn5} and trainability \cite{qnntrain1, qnntrain2, qnntrain3, qnntrain4, qnntrain5} of QML models. Beyond the learnability and trainability issues, QML in practice also presents a considerable hurdle. In scenarios where QML is solely utilized, gate angle encoding is a prevalent technique for input data processing. However, it becomes apparent that scaling issues arise; with increasing input data size, both the width and depth of the quantum circuit must expand accordingly, which can compromise accuracy in the noisy intermediate-scale quantum (NISQ) era \cite{chen2022complexity, chia2024oracle}, massively impacting its practicality. Classical preprocessing techniques might be helpful by decreasing the data's dimensions at the risk of potential loss of crucial information. For QML to be practical, the data dimensions it handles should align with those managed by classical neural networks (NNs). Therefore a better way of data encoding is in demand for a pure QML situation.

Furthermore, a majority drawback for both purely quantum and hybrid quantum-classical machine learning (QCML) \cite{qcml1, qcml2, qcml3} models is the necessity of quantum computing access for the deployment of trained models during the inference phase. Given that quantum computing resources, including cloud-based quantum services, are exceedingly scarce, this requirement poses a substantial barrier to the widespread application of QML models.

To address concerns about practicality, one effective method involves using quantum algorithms to train classical NNs. This approach allows the deployment of the trained classical NNs without facing issues related to data encoding and dependence on quantum computers. Previous research in this area has used quantum walk to search for parameters in classical NNs \cite{tcnnqc1}. However, a Grover-like method in this scenario results in an exponential increase in the circuit depth while scaling.

In this work, we present an innovative QML framework that not only addresses the practicality problem with QML and the dependence on quantum computing resources during the inference stage but also an arising problem with ML regarding the ever-growing parameter size. To be specific, over recent decades, the size of machine learning models has seen a remarkable expansion. A noticeable exponential growth in parameter size since 2018 highlights an increase of five orders of magnitude in just four years \cite{mlpara}. In the foreseeable future, this trend will continue and accelerate, eventually reaching a bottleneck where the gigantic model is too difficult to train with available hardware. In light of this issue, our QML framework provides a significant reduction of trainable parameters down to $O(\text{polylog}(M))$ where $M$ is the required number of parameters for the classical NNs. This framework consists of mapping the weights of a classical NN into the Hilbert space associated with the QNN’s states. Within this framework, by tuning the parameterized circuit (QNN)’s parameters, we effectively manipulate the probability distribution of the quantum states, thereby making adjustments to the parameters of classical NNs. This approach does not require an effective classical data embedding method to quantum form as the data is fed into the classical NN’s as inputs. Moreover, once the model is trained, it is completely independent from quantum computers. This greatly lowers the barrier of using the results of QML. 

In Sec.~\ref{sec:method}, we delve into the process of mapping quantum states onto classical NNs, elaborating on the construction of QNNs and outlining the training procedures involved. Sec.~\ref{sec:nrd} is dedicated to examining the efficiency of the model, its generalization errors, and its practicality. Finally, in Sec.~\ref{sec:conclusion}, we highlight the promising potential of QT in simplifying model parameters and enhancing accuracy, which paves the way for its broad application and future investigative avenues.

\subsection{Main Results} 
Our main results of the proposed Quantum-Train (QT) can be succinctly summarized as follows:
\begin{itemize}
    \item Addressing data encoding issue for QML: The QT approach eliminates the challenges of data encoding faced by pure QML by using classical data inputs and outputs, thus avoiding the complexities and potential information loss associated with encoding larger datasets into quantum states. This method maintains the computational advantages of quantum mechanics without quantum data encoding that is either hard to scale or subject to information loss.

    \item Model compression during training: Our method significantly reduces the number of parameters needed to train classical NNs with $M$ parameters to just $O(\text{polylog} (M))$. This efficiency is achieved by employing $N = \lceil \log_2 M \rceil$ qubits and a polynomially scaled number of QNN layers, optimizing the training process while minimizing resource usage. 

    \item Hardware requirement for inference: The trained model is designed to operate seamlessly on classical hardware, bypassing the need for quantum computing resources. This feature enhances its applicability, especially considering the currently limited access to quantum computers compared to their classical counterparts.

\end{itemize}

\section{Method}
\label{sec:method}
In this section, we delineate the foundational components of the QT framework, elucidating the association between a classical NN and a quantum state, and describe the parameterization and training of this quantum state.
\subsection{Mapping Quantum State to Classical Neural Network}

We commence by considering a classical NN characterized by the parameter vector $\vec{\theta}$, expressed as:
\begin{equation}
\vec{\theta} = (\theta_1, \theta_2, \ldots, \theta_M),
\end{equation}
representing a consolidated view of the $M$ parameters spanning the classical NN, abstracting away the layer-specific details. We propose a quantum state $\ket{\psi}$ encoded by $N = \lceil \log_2 M \rceil$ qubits, sufficient to span a Hilbert space of dimension $2^{\lceil \log_2 M \rceil}$, thereby ensuring coverage for all $M$ parameters. The quantum state's measurement probabilities in the computational basis, denoted by $|\bra{i} {\psi} \rangle|^2$, range within the interval $[0,1]$, for $i \in \{1,2, \hdots, 2^N\}$. The primary objective of this endeavour is to forge a linkage between these quantum measurement probabilities and the classical NN parameters. Specifically, we aim to derive the parameter $\theta_i$, which conventionally spans the continuum from $-\infty$ to $\infty$, from the quantum probabilities $|\bra{i} {\psi} \rangle|^2$, inherently bounded between 0 and 1. 

In this study, we introduce a mapping model labelled as $G_{\vec{\gamma}}$, which is constructed upon an additional NN with parameters $\vec{\gamma}$. This model's inputs merge the basis information in binary form with the corresponding measured probabilities. An example of an input vector $\vec{x}_i$ for 7-qubit case might be:
\begin{equation}
    \vec{x}_i = [0,1,0,0,1,0,0,0.023],
\end{equation}
which encapsulates the measurement probability $|\langle 0100100| \psi \rangle|^2 = 0.023$. The function $G_{\vec{\gamma}}(\vec{x}_i)$ then maps this input to a specific parameter $\theta_i$ within the classical NN we are targeting, such that: 
\begin{equation}
    G_{\vec{\gamma}}(\vec{x_i}) = \theta_i. 
\end{equation}
It's noteworthy to mention that the input to the mapping model has a length of $N+1$, which implies that the size of $\vec{\gamma}$ could vary within $\text{poly}(N)$, given a polynomial number of layers of mapping NN. By employing the mapping model $G_{\vec{\gamma}}$, the parameters of the classical NN $\vec{\theta}$ are derived for the initial $M$ bases from a total of $2^N$ combinations. The training process for obtaining $G_{\vec{\gamma}}$ will be elaborated on later. Contrary to the mapping approach in prior work \cite{tcnnqc2}, where the sign of a specific parameter index remained constant, the NN mapping model proposed in this study will map the signs dynamically. This flexibility allows the mapping to vary according to different ML tasks, thereby enhancing its applicability across a diverse range of scenarios.

\subsection{Quantum Neural Networks}

We have previously defined the qubit size of the quantum state $|\psi \rangle$ and elucidated how the measurement probabilities correlate with the parameters of the classical NN. Expanding upon this foundation, we now delve into the process of constructing $|\psi \rangle$. This begins with its parametrization, marked by the introduction of rotational gate angle dependency, denoted as $|\psi(\vec{\phi}) \rangle$.

\begin{figure*}[ht]
\centering
\includegraphics[scale=0.21]{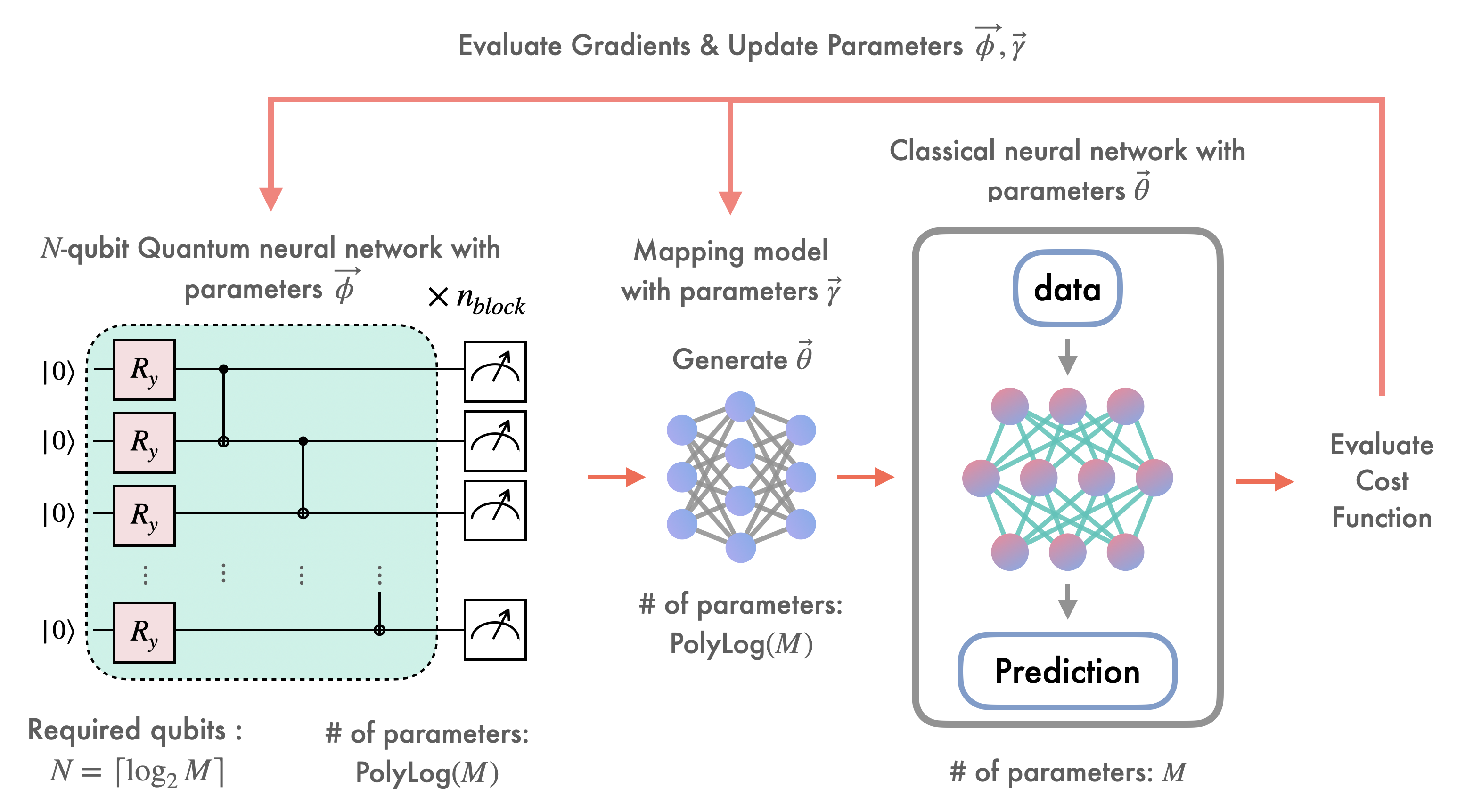}
\caption{QT framework, illustrating the hybrid training process that leverages quantum computing to optimize the parameters of a classical NN. At the left is an $N$-qubit QNN with parameterized $R_y$ and CNOT gates, which are tuned to minimize a cost function, typically the Cross-entropy loss for classification tasks. The QNN's and mapping model $G_{\vec{\gamma}}$'s parameters $\vec{\phi}$ and $\vec{\gamma}$ are updated based on gradient evaluations and are used to generate a corresponding set of classical NN parameters. This model significantly reduces the number of required parameters to $O(\text{polylog}(M))$. The classical NN then processes input data to provide predictions, completing the QT cycle. This framework allows for classical inference post-training, highlighting the QT approach's practicality in bridging quantum and classical computing.}
\label{fig:scheme}
\end{figure*}

Such a parameterized quantum state is pivotal, serving as the QNN \cite{pqcml1}. The construction of this QNN hinges on a chosen ansatz, which lays out the quantum circuit. In our setup, we utilize the quantum gate known as the $R_y$ gate. This gate is crucial for our purposes because it allows us to adjust the quantum state in precise ways with real amplitudes, characterized by its matrix representation:
\begin{equation}
    R_y(\mu) = \left[ \begin{array}{cc}
    \cos(\mu/2) & -\sin(\mu/2) \\
    \sin(\mu/2) & \cos(\mu/2)
    \end{array} \right]
\end{equation}
In conjunction with the $R_y$ gate, the CNOT gate plays a crucial role to entangle qubits.
These parameterized gates, particularly the CNOT with its linear layout, enable the specification of the number of parameters as a polynomial function of the qubit count. A visual depiction of our QNN ansatz is provided in Fig.~\ref{fig:scheme}, which, when paired with the predetermined qubit count, stipulates that the requisite number of parameters scales as $O(\text{polylog}(M))$.

\subsection{Training Flow for Quantum-Train}
After introducing the important elements in the previous sections, the proposed QT framework, and its training flow is described as below. The QT process begins with the establishment of an $N$-qubit QNN, This QNN comprises parameterized $R_y$ gates strategically arranged in blocks. Each block is capable of being repeated $n_{block}$ times, enhancing the model's capacity to encapsulate complex mappings between the quantum states and classical parameters. The QNN parameters, denoted as $\vec{\phi}$, are integral to the generation of classical NN weights through a mapping model characterized by parameters $\vec{\gamma}$.

The selection of activation functions like the hyperbolic tangent (tanh) plays a crucial role in enhancing NNs' ability to model complex data patterns, owing to its non-linear nature and unbounded output range. This characteristic is particularly beneficial in adhering to the principles of the universal approximation theorem \cite{uniapprox1}. Conversely, functions like ReLU, which produce only non-negative outputs, are not utilized in this context. Hence, tanh is employed in the initial layer of the mapping model to foster both theoretical robustness and practical effectiveness in ML flow.

Concurrent to the QNN and mapping model part, the classical NN utilizes the parameter $\vec{\theta}$, generated through the quantum-classical mapping, to undergo traditional training processes. It ingests data, processes it through its architecture, and outputs predictions. The effectiveness of these predictions is quantitatively assessed by the cost function, which provides a measure of performance and directs further optimization. In this work, we consider the classical NN for the classification tasks, thus the cost function utilized is the Cross-entropy loss
\begin{equation}
\label{eq:celoss}
    \ell_{CE} = -\frac{1}{N_{\text{d}}}\sum_{n=1}^{N_{\text{d}}}\left[y_n \log \hat{y}_n + (1-y_n)\log(1-\hat{y}_n) \right]
\end{equation}
with $y_n$ as the true label and $\hat{y}_n$ as the predicted label by the classical NN model for the sample $n$. $N_{\text{d}}$ is the total number of training data. Aim to minimize the loss function Eq.~(\ref{eq:celoss}), the optimization process involves the computation of gradients to both the QNN parameters  $\vec{\phi}$ and the mapping model parameters $\vec{\gamma}$. In the numerical simulation of this work, with state vector simulation method of quantum state, gradients of both the QNN parameter $\vec{\phi}$ and mapping model parameter $\vec{\gamma}$ are calculated analytically as the conventional ML methods. Note that, for shot-based simulations or real quantum devices with a finite number of shots, the gradient of the QNN could be obtained by the parameter-shift rule \cite{ps1, ps2}.  The resulting gradients provides the subsequent updates, which guides the subsequent iterations of parameter adjustments for $\vec{\phi}$ and $\vec{\gamma}$, fostering a feedback loop that progressively hones the classical NN's parameter towards optimal performance.

As highlighted in the flowchart Fig.~\ref{fig:scheme}, the QT framework is engineered for efficiency, necessitating a reduced number of QNN parameters that scale as $O(\text{polylog}(M))$, as opposed to the classical NN's $M$ parameters.  Moreover, the QT approach allows for the trained model to be utilized on classical computers for inference tasks, significantly enhancing the framework's practicality in the context of limited quantum computing resources.


\begin{figure*}[ht]
\centering
\includegraphics[scale=0.25]{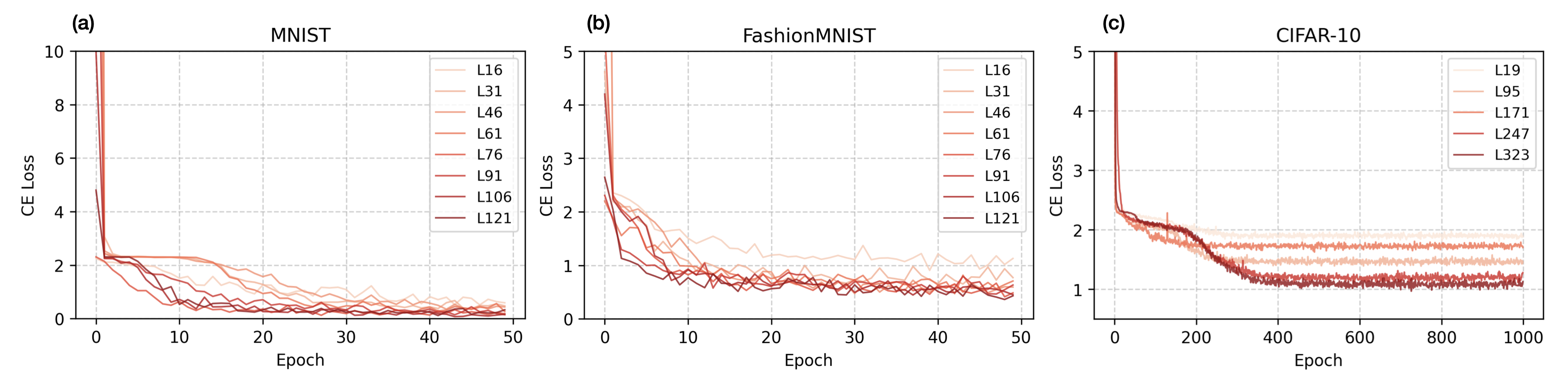}
\caption{Training progression of QT model as visualized through the Cross-Entropy (CE) Loss across epochs for three different datasets: (a) MNIST, (b) FashionMNIST, and (c) CIFAR-10. Each subplot corresponds to a dataset and shows multiple curves, each representing the loss trajectory of a QNN with a varying number of blocks, indicated by ``L'' followed by the block count. For MNIST and FashionMNIST, the CE Loss is plotted over 50 epochs, while for the more complex CIFAR-10 dataset, the loss is tracked over a more extended period of 1000 epochs. The graphs illustrate how the increase in the number of QNN blocks affects the rate and stability of convergence, with a clear trend showing more blocks leading to a steadier decline in loss, indicative of learning improvement.}
\label{fig:result_loss}
\end{figure*}

\begin{figure*}[ht]
\centering
\includegraphics[scale=0.25]{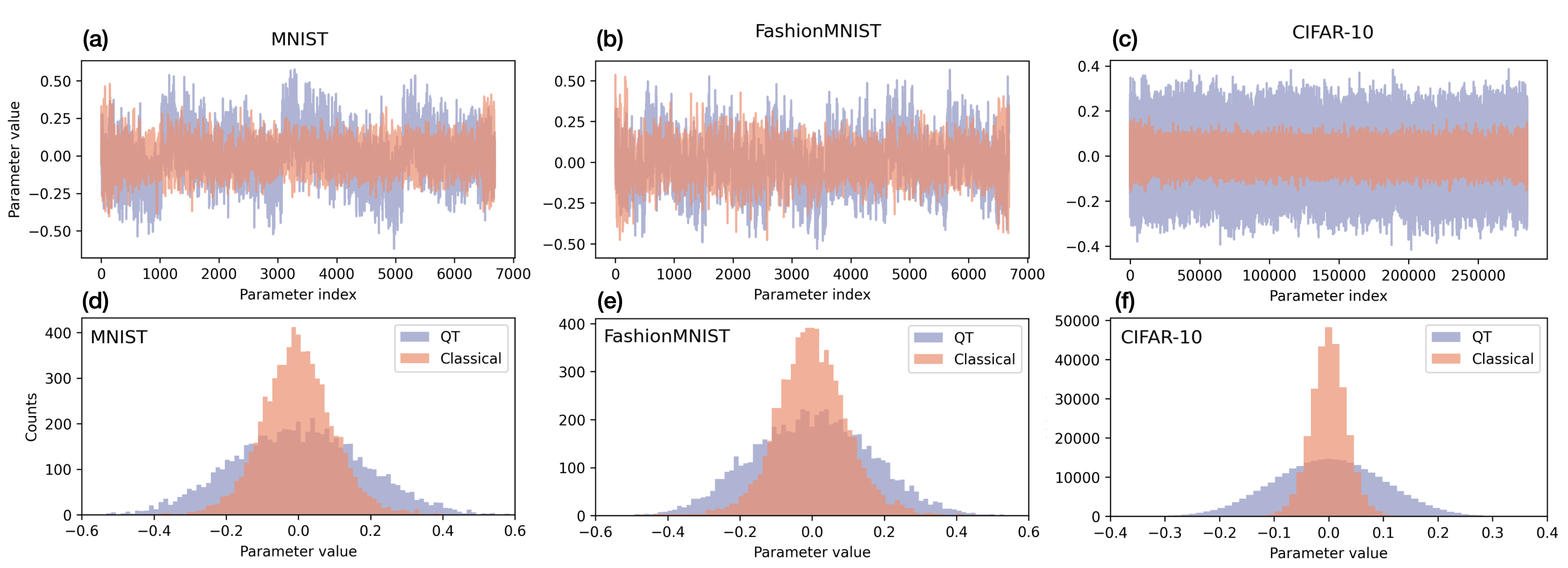}
\caption{Parameter values of classical NN models trained using the QT framework and traditional classical training methods. The top row shows the parameters' values plotted against their indices for the (a) MNIST, (b) FashionMNIST, and (c) CIFAR-10 datasets, respectively, with the QT results corresponding to a QNN with 16 blocks for MNIST and FashionMNIST, and 95 blocks for CIFAR-10. The bottom row (d-f) depicts the distribution of these parameters. The blue color represents the QT-trained model parameters, and the orange color represents the parameters from a classically trained model. The QT approach, which employs quantum computing principles for training, demonstrates a distinctive parameter distribution. }
\label{fig:result_weight_dist}
\end{figure*}

\section{Numerical Results and Discussion}
\label{sec:nrd}
In this section, we elucidate the setup of our numerical simulations and subsequently detail the training outcomes, centering on three pivotal themes: Model Complexity and Efficiency, Generalization Performance, and the Practicality of QT.

Our numerical simulations are conducted using the TorchQuantum package \cite{torchquantum}. The state vector simulation method is employed, beginning with the MNIST dataset \cite{mnist} for a convolutional neural network (CNN) comprising $M = 6690$ parameters, which establishes our classical NN baseline, the input is black and white image of size $28 \times 28$. The requisite qubit count is $\lceil \log_2(6690) \rceil = 13$. The training set includes $60,000$ samples, while the testing set consists of $10,000$ samples. We initialize both the QNN parameters, $\vec{\phi}$, and the mapping model parameters, $\vec{\gamma}$, randomly. The QT model undergoes training for $50$ epochs, utilizing the Adam optimizer \cite{adam} with a batch size of $128$ and a learning rate of $10^{-4}$. The architecture of the classical NN and the QT model, for a varying number of QNN blocks: 16, 31, 46, 61, 76, 91, 106, and 121, is detailed in Table~\ref{table:mnist_res}. This table also includes the test accuracy and parameter size for the resulting model. Fig.~\ref{fig:result_loss}(a) depicts the training loss for different QNN block counts. Identical settings and corresponding results for the FashionMNIST dataset \cite{fashionmnist} are recorded in Table~\ref{table:fashionmnist_res} and Fig.~\ref{fig:result_loss}(b).

In our simulations for CIFAR-10 dataset \cite{cifar10}, we use a CNN with 285,226 parameters, reflecting the dataset's complexity, since the input is colour image of size $32 \times 32$. The model is trained over 1000 epochs with a batch size of 1,000 and a learning rate of \(10^{-4}\). We utilize $\lceil \log_2(285226) \rceil = 19$ qubits to accommodate the model size, investigating the impact of varying QNN block numbers: 19, 95, 171, 247, and 323. This extensive training, with a substantial quantum circuit complexity, aims to rigorously test the QT framework's efficacy on intricate datasets. Model topology and corresponding results are recorded in Table~\ref{table:cifar10_res} and Fig.~\ref{fig:result_loss}(c).

Fig.~\ref{fig:result_weight_dist} examines the parameterization of ML models trained via the QT framework versus classical methods. This figure displays the parameters' values and distributions for MNIST, FashionMNIST, and CIFAR-10 datasets. For MNIST and FashionMNIST, QT employs models with 16 QNN blocks, and for CIFAR-10, 95 QNN blocks are used. Remarkably, the QT models, with their reduced parameter count as shown in Tables ~\ref{table:mnist_res}, ~\ref{table:fashionmnist_res}, and ~\ref{table:cifar10_res}, manage to attain testing accuracies on par with their classical counterparts. Yet, there is a notable divergence in the parameter values and distributions between the two training methodologies, where QT models generally exhibit larger parameter magnitudes and a sparser distribution near zero. This pattern, emerging from training size-$M$ models with only $O(\text{polylog}(M))$ parameters, reflects the unique characteristics of QT model optimization and suggests a fundamentally different parameter efficiency in QT.

\begin{table*}[h!]
\centering
\begin{tabular}{|l|l|c|c|}
\hline
\textbf{Model} & \textbf{Topology} & \textbf{Test acc. (\%)} & \textbf{Para. size} \\
\hline
Classical CNN & (Conv\_layers, max\_pooling)*2, 192-20, 20-10 & 98.45 & 6690 \\
\hline
QT-16 & QNN: ($R_y$ block)*16 + Mapping Model: (14-4, 4-20, 20-4, 4-1) & 81.08 & 457 \\
QT-31 & QNN: ($R_y$ block)*31 + Mapping Model: (14-4, 4-20, 20-4, 4-1)  & 84.69 & 652 \\
QT-46 & QNN: ($R_y$ block)*46 + Mapping Model: (14-4, 4-20, 20-4, 4-1)  & 87.78 & 847 \\
QT-61 & QNN: ($R_y$ block)*61 + Mapping Model: (14-4, 4-20, 20-4, 4-1) & 88.43 & 1042 \\
QT-76 & QNN: ($R_y$ block)*76 + Mapping Model: (14-4, 4-20, 20-4, 4-1) & 90.05 & 1237 \\
QT-91 & QNN: ($R_y$ block)*91 + Mapping Model: (14-4, 4-20, 20-4, 4-1)  & 92.41 & 1432 \\
QT-106 & QNN: ($R_y$ block)*106 + Mapping Model: (14-4, 4-20, 20-4, 4-1)  & 93.81 & 1627 \\
QT-121 & QNN: ($R_y$ block)*121 + Mapping Model: (14-4, 4-20, 20-4, 4-1)  & 93.05 & 1822 \\
\hline
\end{tabular}
\caption{Comparative performance of Classical CNN and QT models with varying numbers of QNN blocks for MNIST dataset.}
\label{table:mnist_res}
\end{table*}

\begin{table*}[h!]
\centering
\begin{tabular}{|l|l|c|c|}
\hline
\textbf{Model} & \textbf{Topology} & \textbf{Test acc. (\%)} & \textbf{Para. size} \\
\hline
Classical CNN & (Conv\_layers, max\_pooling)*2, 192-20, 20-10 & 88.14 & 6690 \\
\hline
QT-16 & QNN: ($R_y$ block)*16 + Mapping Model: (14-4, 4-20, 20-4, 4-1) & 65.69 & 457 \\
QT-31 & QNN: ($R_y$ block)*31 + Mapping Model: (14-4, 4-20, 20-4, 4-1)  & 73.48 & 652 \\
QT-46 & QNN: ($R_y$ block)*46 + Mapping Model: (14-4, 4-20, 20-4, 4-1)  & 76.05 & 847 \\
QT-61 & QNN: ($R_y$ block)*61 + Mapping Model: (14-4, 4-20, 20-4, 4-1) & 74.92 & 1042 \\
QT-76 & QNN: ($R_y$ block)*76 + Mapping Model: (14-4, 4-20, 20-4, 4-1) & 77.31 & 1237 \\
QT-91 & QNN: ($R_y$ block)*91 + Mapping Model: (14-4, 4-20, 20-4, 4-1)  & 77.51 & 1432 \\
QT-106 & QNN: ($R_y$ block)*106 + Mapping Model: (14-4, 4-20, 20-4, 4-1)  & 80.20 & 1627 \\
QT-121 & QNN: ($R_y$ block)*121 + Mapping Model: (14-4, 4-20, 20-4, 4-1)  & 80.74 & 1822 \\
\hline
\end{tabular}
\caption{Comparative performance of Classical CNN and QT models with varying numbers of QNN blocks for FashionMNIST dataset.}
\label{table:fashionmnist_res}
\end{table*}

\begin{table*}[h!]
\centering
\begin{tabular}{|l|l|c|c|}
\hline
\textbf{Model} & \textbf{Topology} & \textbf{Test acc. (\%)} & \textbf{Para. size} \\
\hline
Classical CNN & (Conv\_layers, max\_pooling)*2, 2048-128, 128-64, 64-10 & 62.50 & 285226 \\
\hline
QT-19 & QNN: ($R_y$ block)*19 + Mapping Model: (20-40, 40-200, 200-40, 40-1) & 28.73 & 17482 \\
QT-95 & QNN: ($R_y$ block)*95 + Mapping Model: (20-40, 40-200, 200-40, 40-1) & 44.76 & 18926 \\
QT-171 & QNN: ($R_y$ block)*171 + Mapping Model: (20-40, 40-200, 200-40, 40-1) & 47.42 & 20370 \\
QT-247 & QNN: ($R_y$ block)*247 + Mapping Model: (20-40, 40-200, 200-40, 40-1) & 56.45 & 21814 \\
QT-323 & QNN: ($R_y$ block)*323 + Mapping Model: (20-40, 40-200, 200-40, 40-1) & 60.69 & 23258 \\
\hline
\end{tabular}
\caption{Comparative performance of Classical CNN and QT models with varying numbers of QNN blocks for CIFAR-10 dataset.}
\label{table:cifar10_res}
\end{table*}

\begin{figure*}[ht]
\centering
\includegraphics[scale=0.22]{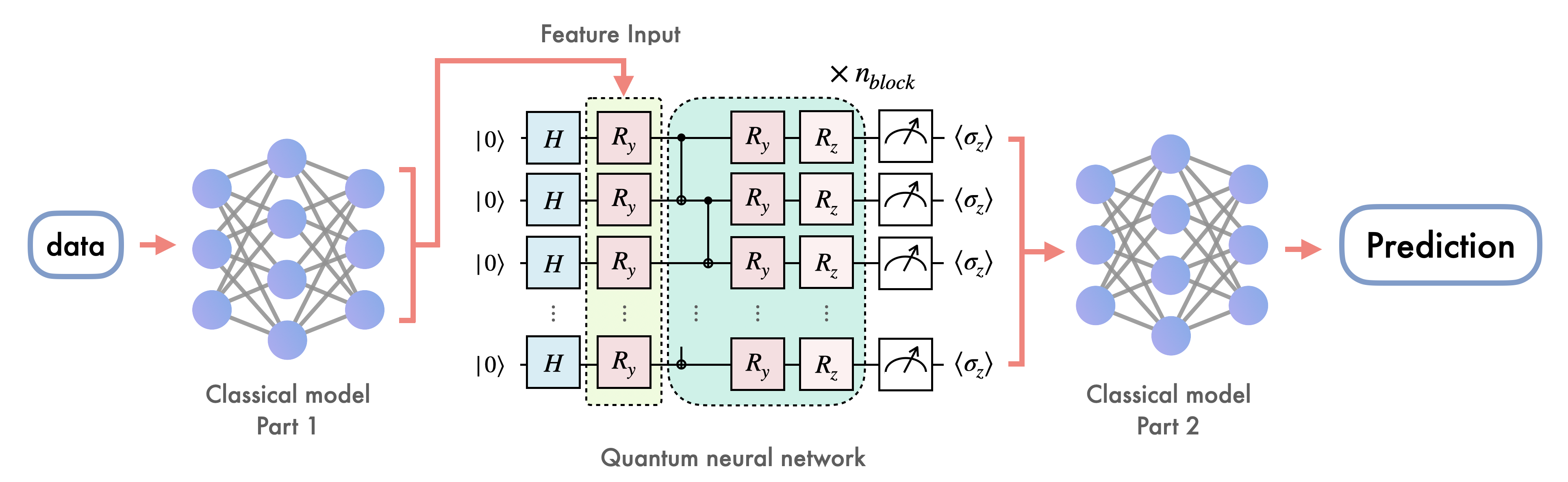}
\caption{Illustration of a Hybrid Quantum-Classical Machine Learning (QCML) model used in the comparison study. The process begins with classical data input, which is first processed by a classical NN (Part 1). The processed features are then fed into a QNN, which consists of multiple qubits initially set to the $|0\rangle$ state. The qubits undergo Hadamard ($H$) transformations to create superposition states, followed by a series of rotation gates $R_y$ and $R_z$ that apply parameterized rotations based on the input features. These rotations encode the data into the quantum state. The QNN may consist of multiple blocks, each contributing to the depth of the quantum circuit. After passing through the QNN, the expectation values $\langle \sigma_z \rangle$ of the qubits are measured, effectively capturing the quantum-processed features. Finally, these features are passed to the second part of the classical NN, which produces the final prediction output. This QCML model leverages the strengths of both quantum and classical computation, aiming to achieve higher model performance and efficiency.
}
\label{fig:qcml_model}
\end{figure*}

Before the further investigation on the QT method, to compare with the existing approach that also use both the classical NN model and QNN during training, here we introduce the QCML \cite{qcml1, qcml2} used in the comparison study later, as shown in Fig.~\ref{fig:qcml_model}. The QCML model integrates classical NN(s) with the QNN, synergizing the computational strengths of both classical and quantum realms. In the QCML architecture, the classical NN first processes the input data, extracting features which are then encoded into a quantum state through a sequence of quantum gates within the QNN. The QNN's architecture is designed with multiple layers, or ``blocks'', enhancing its ability to capture complex patterns in the data. Following the quantum processing, the qubits' expectation values are measured, retrieving information that has been enriched by quantum computation. These quantum-enhanced features are subsequently fed into the latter segment of the classical NN, culminating in the output predictions. This hybrid approach is poised to offer superior performance and efficiency, potentially surpassing the capabilities of models employing solely classical or quantum techniques.

In our setting, specifically for the MNIST dataset, the classical model consists of two parts. Part 1, as illustrated in Fig.~\ref{fig:qcml_model}, employs a CNN with 4693 parameters. The QNN in the middle utilizes 13 qubits across 13 blocks, totaling 169 parameters. Part 2 of the classical model adds an additional 140 parameters, resulting in a total of 5002 parameters for the QCML model. For the FashionMNIST dataset, the setting mirrors that of the MNIST dataset, as these two datasets share the same shapes. In the case of the CIFAR-10 dataset, Part 1 of the classical model is a CNN with 145977 parameters. Similarly, the QNN in the middle utilizes 13 qubits across 13 layers, amounting to 169 parameters. Part 2 of the classical model adds 140 parameters, bringing the total parameters of the QCML model to 146286. It is notable that the majority of parameters in these QCML models originate from the classical parts, with a slight reduction in the total number of parameters. This characteristic results in performance that closely resembles that of the pure classical model, as elaborated on in the next section.

\begin{figure*}[ht]
\centering
\includegraphics[scale=0.25]{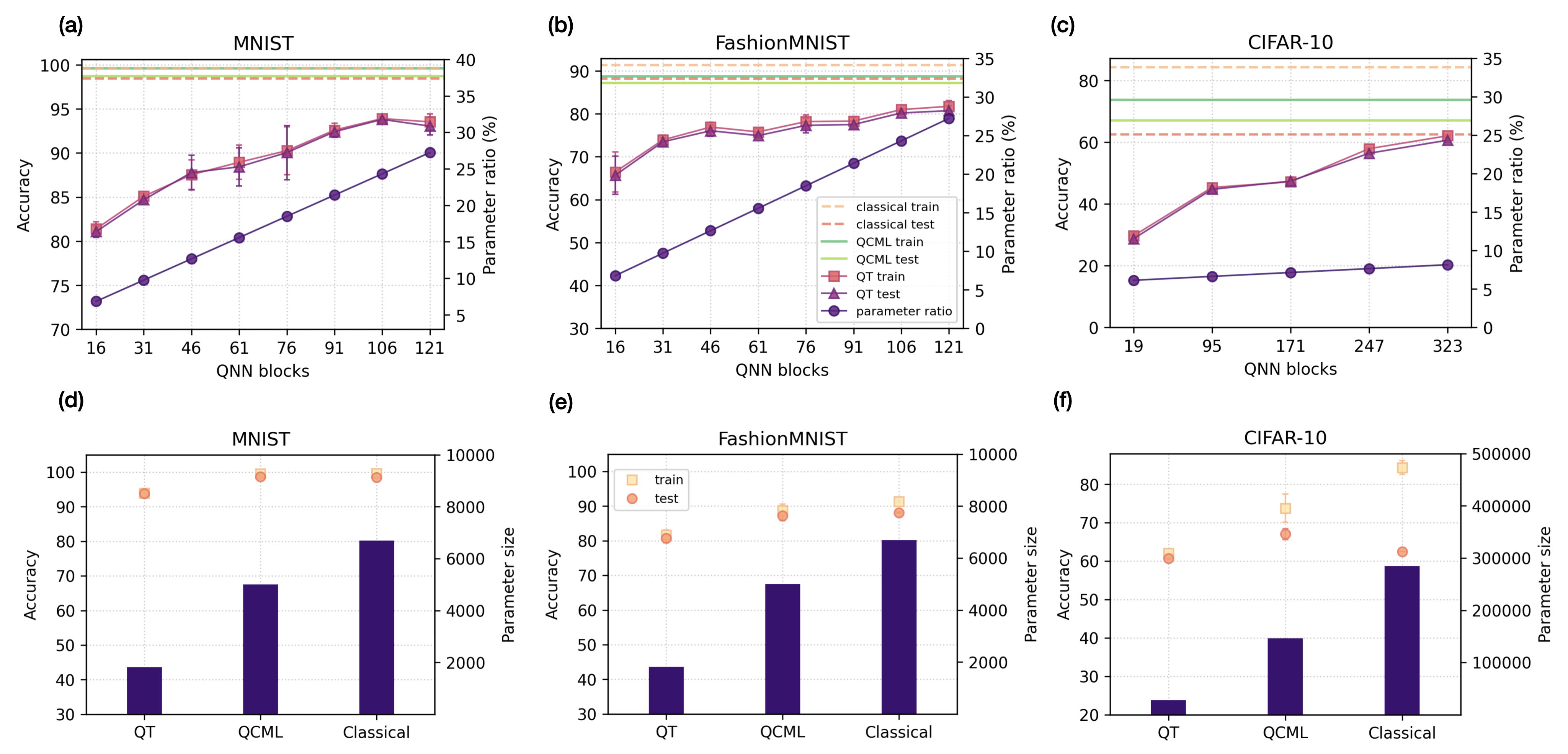}
\caption{(a-c) Performance and parameter efficiency of ML models trained using QT and classical methods across three datasets: MNIST, FashionMNIST, and CIFAR-10. In the top row, the accuracy of QT models increases with the number of QNN blocks used during training, showcasing the potential of QT to achieve competitive accuracy with fewer parameters, as indicated by the parameter ratio line. The bottom row (d-f) contrasts the accuracy and model size (number of parameters) of QT, QCML, and classical training methods, with the QT results corresponding to a QNN with 106 blocks for MNIST and 121 blocks for FashionMNIST, and 323 blocks for CIFAR-10. Notably, QT and QCML methods demonstrate a significant reduction in parameter size for all cases while maintaining high accuracy, suggesting that quantum methods can lead to more compact models.
}
\label{fig:result_para}
\end{figure*}

\begin{figure*}[ht]
\centering
\includegraphics[scale=0.26]{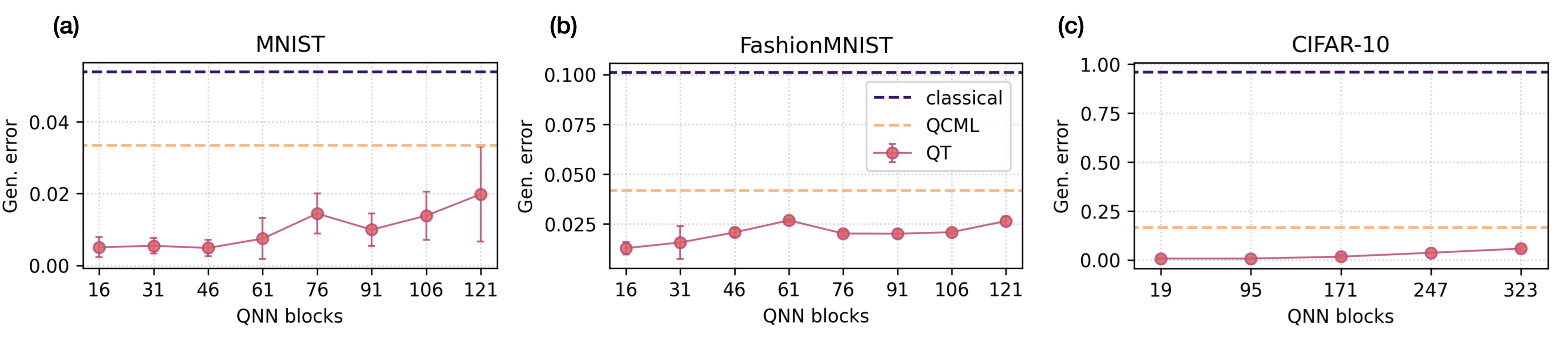}
\caption{Generalization error trends for ML models trained on the MNIST, FashionMNIST, and CIFAR-10 datasets using different training approaches: QT, classical, and QCML. For (a) MNIST, the generalization error of QT models typically decreases as the number of QNN blocks increases, which suggests that model performance improves with greater quantum model complexity. In contrast, for (b) FashionMNIST, the generalization error of QT models increases with more QNN blocks but remains lower than that of QCML and classical approaches. For (c) CIFAR-10, the generalization error is relatively low and stable regardless of the number of QNN blocks.
}

\label{fig:result_gen_error}
\end{figure*}

\subsection{Model Complexity and Efficiency}
\label{sec:model_complexity_efficiency}

The results of QT demonstrates that increasing the number of QNN blocks enhances accuracy across MNIST, FashionMNIST, and CIFAR-10 datasets, as shown in Table~\ref{table:mnist_res}, \ref{table:fashionmnist_res}, \ref{table:cifar10_res}, and Fig.~\ref{fig:result_para}(a-c). Specifically, for MNIST, QT achieves 93.81\% testing accuracy with 24.3\% of the classical approach's parameters, where the testing accuracy of the classical approach is 98.45\%. For FashionMNIST, QT attains a testing accuracy of 80.74\% compared to the 88.14\% of the classical model while using 27.2\% of parameters. Remarkably, in CIFAR-10, QT reaches a testing accuracy of 60.69\% with just 8.1\% of the parameters used by the classical model, where the testing accuracy of classical approach is 62.50\%, showcasing remarkable parameter efficiency with slightly compromising on testing accuracy. 

In comparing the QT with purely classical and QCML approach, we observe distinct performance patterns. Fig.~\ref{fig:result_para}(d-f) delineates the training and testing accuracies, alongside the parameter sizes, for QT, QCML, and classical approaches, with QT configurations employing L16 blocks for MNIST and FashionMNIST, and L95 blocks for CIFAR-10. Notably, QCML outperforms the others in MNIST and FashionMNIST, achieving optimal results with a moderate parameter count, a trend we will further dissect in Sec.~\ref{sec:prac}. For CIFAR-10, QT and QCML show comparable performance, with a slight edge for QT in training accuracy and for QCML in testing accuracy. Interestingly, the classical approach, despite superior training accuracy, exhibits the lowest testing accuracy, implying a higher generalization error, a topic we'll explore in subsequent sections. This evaluation highlights the nuanced trade-offs between these methods and sets the stage for a deeper exploration of their practical applications.

\subsection{On Generalization Error}
Reflecting on the results depicted in Fig.\ref{fig:result_para}(a-c), we note a consistent pattern: the classical approach exhibits a wider gap between training and testing accuracies compared to QT, indicative of a larger ``generalization error''\cite{gen_error_1}. Generalization error, a term referring to a model's performance on new, unseen data, is crucial in evaluating machine learning efficacy. As defined by 
\begin{equation}
    \text{gen}(\alpha) = R(\alpha) - R_S(\alpha),
\end{equation}
where the learnable parameter $\alpha$, with the expected loss 
\begin{equation}
    R(\alpha) = \mathbb{E}_{(x,y)\sim P}[\ell(\alpha; x,y)]
\end{equation}
of data input $x$ and output $y$ over the data-generating distribution $P$, and 
\begin{equation}
    R_S(\alpha) = \frac{1}{N_{\text{d}}} \sum_{n=1}^{N_{\text{d}}} \ell(\alpha; x_n, y_n)
\end{equation}
is the average loss over the training set. 

Fig.~\ref{fig:result_gen_error} demonstrates the generalization errors across various models, accentuating the classical models' tendency for overfitting compared to QT models. The QT framework, exploiting quantum mechanics, seems to endow its models with superior generalization capabilities over traditional data. This characteristic is shared by QCML models, which exhibit similar generalization errors to QT. Commonly, one might anticipate that models with more parameters would overfit training data, yielding higher generalization errors. However, an increase in training data typically results in a better estimate of the true loss and, consequently, a lower generalization error. This phenomenon underscores the intricate balance between model complexity and the volume of training data in achieving optimal model generalization.

As the number of QNN blocks and consequently, model complexity increases within the QT framework, the generalization error for the MNIST dataset similarly rises. The FashionMNIST dataset also shows increased generalization error with a higher count of QNN blocks, yet this error remains below that of the classical and QCML models, same as the MNIST case. CIFAR-10's QT models maintain low and stable generalization errors, with only a slight increase observed at 247 QNN blocks. This stability underscores QT's capability to effectively handle the complexities of diverse datasets. This pattern aligns with findings from existing research \cite{gen_error_1}, suggesting that the generalization error is proportional to QNN parameter size with roughly $\text{gen}(\alpha) \sim \mathcal{O}(\sqrt{N_t/n})$, where $N_t$ is the number of parameters and $n$ is the number of training samples. It is important to note that the focus here is on the trend of proportionality rather than on a precise scaling function, providing a broader perspective on how QNN complexity impacts model generalization.

The previous work \cite{gen_error_2} illustrates the concept of over-parameterization, where over-parameterization occurs when the number of parameters $N_t$ significantly exceeds the number of training samples $n$, $N_t \gg n$. For QT models, considered in the category of the QNN, the threshold for over-parameterization is lower, indicated simply when $N_t > n$. In our study, neither classical, QCML, nor QT models for MNIST and FashionMNIST meet these conditions, given $n = 6 \times 10^4$ and parameter sizes on the order of $10^3$. For CIFAR-10, the classical model is over-parameterized with 285,226 parameters against 50,000 training samples. In contrast, QT models, with around 23,000 parameters, avoid over-parameterization, contributing to their generalization performance.

\subsection{Practicality of Quantum-Train}
\label{sec:prac}

Over recent decades, the complexity of machine learning models has dramatically increased, marked by an exponential surge in the number of parameters. This trajectory is expected to persist and even accelerate, eventually leading to a critical point where training these colossal models becomes impractical with existing hardware capabilities, or scaling them linearly becomes prohibitively expensive (e.g., doubling the number of GPUs does not consistently yield a twofold increase in speed). Conversely, the QT approach, which employs a polylogarithmic number of parameters for training classical machine learning models, could signify a pivotal shift. It potentially moves us away from linear hardware scaling requirements for training large models, with the power of quantum computing. 

In pure QML, data encoding into quantum states, such as through gate angle encoding, poses a challenge due to the constraints on input data size. More complex data necessitates additional gate angles for encoding, leading to potentially deep quantum circuits for larger datasets. An alternative approach involves preprocessing and compressing the data before encoding it into quantum states, though this risks losing vital information. In contrast, the proposed QT methodology leverages classical inputs and outputs, effectively mapping quantum states to classical NN parameters. This approach retains the computational benefits of Hilbert space while circumventing the drawbacks associated with direct data encoding in quantum circuits.

The QCML model aims to leverage the strengths of classical NNs for data preprocessing and postprocessing while utilizing QNN to process data within Hilbert space, aiming for enhanced performance with fewer parameters. As indicated in Fig.~\ref{fig:result_para} and Fig.~\ref{fig:result_gen_error}, QCML appears to outperform QT and also shows parameter efficiency compared to classical methods. It's noteworthy, however, that both pure QML and QCML models necessitate a quantum computer for executing QNN blocks during inference stage. In contrast, the QT model's training outcome is a purely classical model that can operate independently on classical hardware, bridging the hardware gap typically associated with QML during inference and offering a practical solution for widespread application.

To fully realize the practical potential of the QT method, it's crucial to address its existing challenges. A notable concern arises from the state vector simulation method employed in this study, which operates under the assumption of noise-free conditions and infinite measurement shots to determine probability distributions. While suitable for a proof-of-concept, such conditions are idealized compared to the noise and finite measurements encountered in actual quantum hardware or simulations.

Similar to other research on QNN with multiple layers, an effective error mitigation or correction strategy, or alternatively, the use of high-fidelity qubit systems, may be necessary. Such measures are crucial to maintain quantum coherence up to the final layer of the QNN, ensuring the network fulfills its intended function and rendering the QT approach feasible in practice.

The exact impact of the number of measurement shots on accuracy in realistic scenarios remains to be clarified. However, drawing parallels from theoretical findings in quantum tomography might shed light on expected scaling trends. For example, in fidelity tomography tasks—where the goal is to achieve a fidelity $F(\ket{\psi}, \ket{\phi}) \geq 1 - \epsilon$ between the output state $\ket{\phi}$ and an unknown state $\ket{\psi}$ with a given infidelity $\epsilon$—previous research \cite{haah2016sample} suggests that a sufficient number of measurement shots is given by $N_{\text{shot}} = O(\frac{2^N}{\epsilon} \log(\frac{2^N}{\epsilon}))$. Given the exponential increase in required measurement shots with system size $N$, a linear scaling strategy in relation to $2^N$ appears to be a judicious approach for improving both training and testing accuracies in larger systems. Consequently, the ability to rapidly and frequently acquire measurement outcomes in quantum systems becomes a pivotal factor for the practical implementation and success of the QT approach. Generative models are a potential approach that can be particularly beneficial in this context by efficiently simulating measurement outcomes \cite{gen_qst_1}, thus mitigating the issue of the exponential number of measurement shots required. By leveraging generative models, it is possible to reduce the computational burden and enhance the feasibility of large-scale quantum state tomography, thereby making our QT approach practical.

\section{Conclusion}
\label{sec:conclusion}

In this study, we introduced a groundbreaking framework, QT, that merges quantum computing principles with classical ML algorithms to tackle three key challenges: data encoding in QML, model compression in classical models, and hardware requirements for inference in QML and QCML. The QT framework initiates with the formation of an $N$-qubit QNN, comprising blocks of parameterized gates. The parameters of the QNN, play a crucial role in deriving classical NN parameters through a mapping model.

Our experimental findings reveal that this hybrid approach significantly reduces model complexity and size while preserving or slightly reduces predictive accuracy. These results highlight the transformative potential of quantum computing in ML, especially in scenarios limited by computational resources. Furthermore, this approach contributes to a reduction in the generalization error of ML models due to decreased parameters, thereby bolstering model robustness compared to traditional classical methods.

We demonstrate that QT can effectively reduce the number of parameters within an NN. An intriguing avenue for exploration involves leveraging other aspects of quantum computing tools to enhance the training process of standard LLMs (Large Language Models). For instance, \cite{alman2024fine, hu2024computational, hu2024computationalc} illustrate the lower bound of computing the gradient of the loss function for a one-layer attention network, assuming the Strong Exponential Time Hypothesis (SETH). However, Grover's algorithm suggests the potential for quantum enhancements in training LLMs. On the other hand, the lower bound of Grover's algorithm \cite{boyer1998tight} indicates the similar conjecture called QSETH \cite{buhrman2019quantum} indicates that a similar lower bound of training LLM can be achieved via a quantum circuit.

While our study utilizes classification tasks for ML models as a proof-of-concept, the proposed framework, as illustrated in Fig.~\ref{fig:scheme}, is inherently versatile, applicable to any ML model characterized by $M$ parameters. For instance, in the realm of (quantum) reinforcement learning (RL), implementing QT could decouple the QNN from quantum hardware during the inference stage, offering significant advantages, particularly in RL scenarios. This versatility paves the way for a myriad of new applications, merging existing ML methodologies with practical problems and exploring the effects of QT's parameter reduction. 


\acknowledgments{
H.-S.G. acknowledges support from the National Science and Technology Council, Taiwan under Grants No.~NSTC 113-2119-M-002 -021, No.~NSTC112-2119-M-002-014, No.~NSTC 111-2119-M-002-007, and No.~NSTC 111-2627-M-002-001, from the US Air Force Office of Scientific Research under Award Number FA2386-20-1-4052, and from the National Taiwan University under Grants No.~NTU-CC-112L893404 and No.~NTU-CC-113L891604. H.-S.G. is also grateful for the support from the ``Center for Advanced Computing and Imaging in Biomedicine (NTU-113L900702)'' through The Featured Areas Research Center Program within the framework of the Higher Education Sprout Project by the Ministry of Education (MOE), Taiwan, and the support from the Physics Division, National Center for Theoretical Sciences, Taiwan. J.-G.Y. and Y.-J.C. acknowledge the funding provided under the ITRI EOSL 2022 project``Quantum Circuit Design and Application''. E.-J.K. thanks the National Center for Theoretical Sciences of Taiwan for funding (112-2124-M-002-003)}

\nocite{*}

\bibliography{references}

\end{document}